\documentclass[a4paper,dvips,12pt]{article}
\usepackage{axodraw}
\usepackage{array,tabularx,epsfig,graphicx,rotating}
\usepackage{ifthen}
\usepackage{fancybox}
\usepackage{amsfonts}

\newcommand{\beq}{\begin{equation}}
\newcommand{\eeq}{\end{equation}}

\begin{document}
%\vspace*{8mm}

\begin{flushright}
SINP 2004-5/744\\
January\,, 2004\\
%ZEUS-Note 04--*** \\
%January\,, 2004\\
\end{flushright}  
\begin{center}
{\large 
\bf 
Kinematic selection criteria in a new resonance searches:\\
Application to pentaquark states}

\vspace{5mm} 
{ \bf Boris  Levchenko}\\
{\it Skobeltsyn Institute of Nuclear Physics, Moscow State University }\\

\end{center}

\abstract{
In this note I recall some features of two-body decay kinematics which can be  effectively   applied, in particular, in experimental searches 
for  pentaquark states.}

\section{Introduction}

\vspace*{5mm}

In experimental searches for  resonances, tracks  of secondary particles
are combined to form   resonance candidates. In high energy reactions 
with multiparticle final states, this method of  resonance
reconstruction may cause a huge combinatorial background.
Any 
additional physical information about the resonance  and its     
 decays (life-time, the mass of secondary  particles etc.) 
 helps considerably to reduce the background.
For instance, the presence in an event a well separated secondary vertex
allows us to reconstruct $\Lambda,\, K_s$ and mesons from the $D$-meson  family
with a very low background. One  such example is  in Ref. \cite{ksks}.
Moreover, if  masses of decay  products  are known and significantly
differ from each other,  this information also has to be used to reduce the  background.
This note is a discussion of these issues.

Below  only two-body decays are considered  and 
 for each  final state particle  a set of equations 
describing  boundaries of the physical region  is given.  
Features of each physical region, through implementation in selection
 criteria, can be used in the background suppression.

\section{  Two particle decay}

Let us consider a  two-particle decay, $R\rightarrow {\bf a}+{\bf b}$, 
of a resonance $R$ with a mass $M_R$ and a momentum $\vec{P_R}$ in the laboratory frame. 
Masses of the decay products are denoted as $m_{\bf a}$ 
and $m_{\bf b}$. At the rest frame of  $R$  the particles ${\bf a}$ and ${\bf b}$ 
are flying in  opposite directions with the momentum 
\cite{kopbk}

\beq
P^*\,=\,\frac{1}{2M_R}[ (M^2_R - m^2_{\bf a} - m^2_b)^2 - 4m^2_{\bf a} m^2_{\bf 
b} ] ^{1/2}
\label{eq:1}
\eeq 
In the laboratory frame, the 
absolute momenta $p_{\bf a}$ and $p_{\bf b}$ of the particles  ${\bf a}$ and  ${\bf b}$  
depend on the 
relative  orientation  of the rest frame vectors $\vec{p}^*_{\bf a}$ and 
$\vec{p}^*_{\bf b}$  with respect to the boost vector. 
We shall consider only  Lorentz boosts along the momentum $\vec{P}_R$ 
and denote by $\theta^*_{\bf a}$ the polar angle between $\vec{p}^*_{\bf a}$ and 
the direction  given by $\vec{P}_R$.
In that case,  $\theta^*_{\bf b} = \pi - \theta^*_{\bf a}$.

The energy and momentum components in both frames are related 
via \cite{kopbk}  

\beq 
E_{{\bf a}(b)} = \gamma E^*_{{\bf a}({\bf b})} - \vec{\eta}\cdot\vec{p}^*_{{\bf 
a}(b)}
\label{eq:2}
\eeq
\beq 
\vec{p}_{\bf a(b)} =\vec{p}^*_{\bf a(b)}  + \vec{\eta}\, 
[\frac{\vec{\eta}\cdot\vec{p}^*_{\bf a(b)} }{\gamma +1} - E^*_{\bf a(b)}]
\label{eq:3}
\eeq
For a boost along $\vec{P}_R$,  the boost  parameters are
\beq 
\gamma = \frac{E_R}{M_R}, \ \ \ \ \ \vec{\eta}= -\frac{\vec{P}_R}{M_R}
\label{eq:31}
\eeq
and therefore
\beq
p_{\bf a(b)}=\frac{1}{M_R}\sqrt{(E_RE^*_{\bf a(b)}+P_RP^*cos{\theta^*_{\bf 
a(b)}}  )^2-m^2_{\bf a(b)}M^2_R}
\label{eq:32}
\eeq
If in an experiment there is no possibility  to determine a  particle type
corresponding to a given charged track,  then
only information about the particle momentum, Eq.(\ref{eq:32}), is used. 
In the opposite case,
when the particle type can be identified, for instance, by ionization, 
time of flight etc., one may incorporate this information and in addition
use  Eq. (\ref{eq:2}). Below, these cases are considered separately.

\subsection{ Particles without identification}

The boundaries of the  physical regions of the particle ${\bf a}$ 
on the $(P_R,p_{\bf a})$ plane  are easy to  obtain with the use of the 
equation (\ref{eq:32}). For a given $P_R$, $p_{\bf a}$ reaches
the upper limit at $\theta^*_{\bf a}=0$ 
\beq 
p^+_{\bf a}=\frac{1}{M_R}\left(P_RE^*_{\bf a} + E_RP^* \right)
\label{eq:4}
\eeq
when the lower limit  at $\theta^*_{\bf a}=\pi$
\beq 
p^-_{\bf a}= \frac{1}{M_R}| P_RE^*_{\bf a}  - E_RP^* \,|
\label{eq:5}
\eeq
The equations similar to (\ref{eq:4})-(\ref{eq:5})  are valid  for the 
particle ${\bf b}$ too. At the resonance rest frame
 (\ref{eq:5}) demonstrates an interesting feature 
of the momentum  of the particle flying backward. 
With increasing  $P_R$,
the momentum $p_{\bf a}$ ($p_b$) first decreases,  
at $ P_R = M_R P^*/m_{{\bf a}(b)}$ it 
reaches the zero value and only at lager $P_R$ starts to increase. 
When plotted on the momentum $(P_R,p_{{\bf a}})$ plane,
Eqs.(\ref{eq:4})-(\ref{eq:5})  select a band-like  physical region ($m$-band). 
 For secondary particles with equal masses, $m_{\bf a}=m_{\bf b}$, these $m$-bands 
are fully overlapping. 
However, if for instance, $m_{\bf a}> m_{\bf b}$, the  $m$-bands 
 overlap only partially or even separate out at 
 \beq 
 P_R \geq P^{(s)}_R=\frac{ 2M_RP^* }{ \sqrt{ (E^*_{\bf a}   -  E^*_{\bf  b})^2 - 
4(P^*)^2  }}
 \label{eq:6}
\eeq
 The last equation follows from the condition $ p^-_{\bf a} \geq p^+_{\bf b}$. 
In Eq (\ref{eq:6}) the expression under the square root is positive  only if 
 \beq 
 P^*<\frac{ m^2_{\bf a}   - m^2_{\bf b} }   {  \sqrt{  8( m^2_{\bf a}  + 
m^2_{\bf b}    ) }  }
\label{eq:7}
\eeq
Thus,    for $m_{\bf a} \gg m_{\bf b}$ the physical regions  of the  particle 
${\bf a}$  and ${\bf b}$  do not overlap if 
\beq 
m_{\bf a} \geq \sqrt{8}P^*
\label{eq:8}
\eeq
and $P_R > P^{(s)}_R$.

We shall now consider a few particular applications of equations 
(\ref{eq:4})-(\ref{eq:8}). 
According to the PDG \cite{pdg} only resonances with mass  close to the 
threshold value
$m_{\bf a}+m_{\bf b}$  satisfy the condition (\ref{eq:8}).   Some of them
are listed in Table 1, where $P^*$ and $P_R^{(s)}$ values were calculated with 
use of 
(\ref{eq:1}) and
(\ref{eq:6}), respectively.

\begin{center}
 \hspace*{105mm}{Table 1 }\\
 \vspace*{2mm}
 \begin{tabular}{|l|l|l|l|l|} 
 \hline
  $R\rightarrow {\bf a}+{\bf b}$& $M_R,$ GeV & $m_{\bf a}$, GeV  & 
$\sqrt{8}P^*$, GeV&$P_R^{(s)}$, GeV\\ 
\hline
\hline 
 $\Lambda\rightarrow p+\pi^- $       &1.115  & 0.938 & 0.284  &  0.301 \\ 
\hline
 $\Delta\rightarrow N+\pi $          &1.232  & 0.939 & 0.643  &  1.052 \\ 
\hline
 $\Sigma\rightarrow N+\pi $          &1.193  & 0.939 & 0.528  &  0.714 \\ 
\hline
 $\Sigma\rightarrow \Lambda+\pi $    &1.385  & 1.115 & 0.596  &  0.752 \\ 
\hline
 $\Sigma\rightarrow N+K$             &1.480  & 0.939 & 0.494  &  2.066 \\ 
\hline
 $D^0\rightarrow  a_1(1260) ^+ +K^-$ &1.865  & 1.230 & 0.924  &  6.457 \\ 
\hline
 $D^{* \pm}\rightarrow D^0 +\pi ^{\pm}$   & 2.010 & 1.865 & 0.107 & 0.088 \\ 
\hline
 $D^+_{s1} \rightarrow D^*(2010)^+ + K^0$& 2.535 & 2.010 & 0.424 & 0.519\\ 
\hline
 ...& ...& ...& ...& ...\\ 
\hline
 \end{tabular}
\end{center}
\vspace*{3mm}

Figs. 1a , 1b and 1c  shows $m$-bands of decay products
for some resonances listed in Table 1. For all of them always
\beq 
p_{\bf a} > p_{\bf b} 
\label{eq:9}
\eeq
at $P_R > P^{(s)}_R$, and $m_{\bf a}$ significantly larger than $ m_{\bf b}$. 
We will  refer to the momentum condition (\ref{eq:9}) as a  $m$-selector.
Thus, for resonances of the type  in Table 1, 
fulfillment of the  condition (\ref{eq:9}) allows an  assignment to the particle ${\bf 
a}$ of the mass $m_{\bf a}$, i.e. we identify the particle ${\bf a}$.
The $m$-selector (\ref{eq:9}) is a powerful tool for  background suppression.

%We can also conclude, the resonance reconstruction with the selection 
%(\ref{eq:9}) removes background combinations.

In the two-body  decay modes one meet  with a higher incidence the situation  
shown in Fig.~1d. For these decays the  conditions (\ref{eq:7})-(\ref{eq:8}) are
not fulfilled and at any $P_R$ the phase space bands remain overlapping.
Nevertheless, if $m_{\bf a}> m_{\bf b}$, one can demand 
fulfillment of (\ref{eq:9}) 
for  the particle ${\bf a}$ in order to  assign  it the mass $m_{\bf a}$. 
In course of the resonance
search   the  condition 
(\ref{eq:9}) rejects not only a significant part of the background, but also 
some fraction of the signal combinations. To estimate the efficiency of the 
 $m$-selector we proceed in the following way.

On $(P_R,p_{\bf a})$ plane the condition (\ref{eq:9}) is valid up to the line 
defined by the equation
\beq 
p_{\bf a}(cos{\theta^*_{\bf a}},P_R)=p_{\bf b}(cos{\theta^*_{\bf b}},P_R)
\label{eq:10}
\eeq
For $K^*\rightarrow K\pi$ decays the last equality is shown in Fig.~1d by the dashed line.
The true $K^*$ is a combination of a kaon  from the region above the line 
(\ref{eq:10}) with a pion below it, and vice versa.
From (\ref{eq:10}) and (\ref{eq:32}) we find the variation of 
$\theta^*_{\bf a} = \pi - \theta^*_{\bf b}$ along the line (\ref{eq:10})
\beq 
cos{\hat{\theta}^*_{\bf a}}= -\frac{P_R}{E_R}\cdot\frac{(E^*_{\bf a}-E^*_{\bf 
b})}{2P^*}
\label{eq:11}
\eeq
At the rest frame of  $R$, (\ref{eq:9}) is equivalent to the exclusion of the region  
$\theta^*_{\bf a}>\hat{\theta}^*_{\bf a}$.
For unpolarized  particles the distribution of $cos{\theta^*_{\bf a}}$ is 
uniform and   we define the efficiency of the $m$-selector  (\ref{eq:9}) as
\beq 
E_{ff}=\frac{1-cos{\hat{\theta}^*_{\bf a}}}{2}\cdot 100\%
\label{eq:12}
\eeq
With the definition (\ref{eq:12}) we get $E_{ff}=100\%$,
when  the conditions (\ref{eq:7})-(\ref{eq:8}) are fulfilled, and  $E_{ff}=50\%$ for 
decays with $m_{\bf a}= m_{\bf b}$
\footnote{For real data $E_{ff}<50\%$, see discussion 
in Sec.3}. 
 As another example,  in Fig.~4 by the dashed line is shown the evolution of $E_{ff}$ 
with  $P_R$   in $K^*\rightarrow K\pi$ decays. $E_{ff}$  grows because
with the increase of $P_R$ the overlap of $m$-bands decreases.

In the limit large $P_R$, the expression for the efficiency can be simplified:
\begin{eqnarray}
E_{ff}&=&
\frac{1}{2}\left ( 1 + \frac{m^2_{\bf a}-m^2_{\bf b}}
{\sqrt{\left [M_R^2-(m_{\bf a}-m_{\bf b})^2\right ]\left [M_R^2-(m_{\bf a}+
m_{\bf b})^2\right ]}}\right ) \cdot 100\%  \label{eq:13a} \\
&\simeq&
\frac{1}{2}\cdot\frac{M_R^2}{M_R^2-m^2_{\bf a}}\cdot 100\%
\label{eq:13b}
\end{eqnarray}
The last equation is a good approximation only if $m_{\bf a}\gg m_{\bf b}$.
Equations (\ref{eq:13a})-(\ref{eq:13b}) confirms the result we have already seen in  Fig.~4a,
 where $E_{ff}$ is independent of
$P_R$ at large $P_R$. On the other hand, for fixed values of $m_{\bf a}$ and $m_{\bf b}$, 
 $E_{ff}$  decreases with increasing $M_R$, the mass of the resonance candidate.
Thus, the higher the invariant mass of a two-particle combination,
 the more strongly (\ref{eq:9}) suppresses that part of the  mass spectrum.

\subsection{ Identified particles}

Accounting for the particle masses\footnote{The author is grateful to
S.Chekanov for a discussion of that subject.} transforms Eqs (\ref{eq:4})-(\ref{eq:5}) into
\beq 
E_{\bf a}^+=\frac{1}{M_R}\left(E_{R}E^*_{\bf a}\,+\,P_RP^*\right)
\label{eq:14}
\eeq
\beq 
E_{\bf a}^-=\frac{1}{M_R}\left(E_{R}E^*_{\bf a}\,-\,P_RP^*\right)
\label{eq:15}
\eeq
and  at low $P_R$ leads  to a 'repulsion' between the phase space $E$-bands 
on the energy ($P_R,E_{\bf a}$) plane (Fig. 2).
For all resonances with $m_{\bf a} \gg m_{\bf b}$ (see Table 1)
\beq 
E_{\bf a}>E_{\bf b}
\label{eq:16}
\eeq
independently of $P_R$. 
This is not always true for the background combinations.
Therefore, if applied, the condition (\ref{eq:16}) 
suppresses the background  even more strongly than (\ref{eq:9}).
We will  refer to the energy condition (\ref{eq:16}) as a  $E$-selector.

As in the previous section, if the masses of the secondary particles do not differ 
significantly, at $ P_R$ greater than
\beq 
\hat{P}_R=\frac{M_R(E^*_{\bf a} - E^*_{\bf b} )}{\sqrt{4(P^*)^2-(E^*_{\bf a} - E^*_{\bf b} )^2}}
\label{eq:17}
\eeq
the $E$-bands start overlapping in the way  shown in Fig 2d. The loss of signal combinations
by demanding  (\ref{eq:16}) for resonance candidates we estimate  again with Eq. (\ref{eq:12}). 
In the case under consideration,
 $cos{\hat{\theta}^*_{\bf a}} $ is a solution of the equation
\beq 
E_{\bf a}(cos{\theta^*_{\bf a}},P_R)=E_{\bf b}(cos{\theta^*_{\bf b}},P_R)
\label{eq:18}
\eeq
Thus
\beq 
cos{\hat{\theta}^*_{\bf a}}= -\frac{E_R}{P_R}\cdot\frac{(E^*_{\bf a} -E^*_{\bf b} )}{2P^*}
\label{eq:19}
\eeq
In the limit large $P_R$, with (\ref{eq:19}) we again recover Eqs.~(\ref{eq:13a})-(\ref{eq:13b}).
The evolution of $E_{ff}$  in $K^*\rightarrow K\pi$ decays is shown in Fig.~4a
 by the full line. $E_{ff}=100\%$ at $P_R<\hat{P}_R$ 
and drops down up to the value (\ref{eq:13a}), $E_{ff}\simeq 71\%$, with increase of  
$P_R$.

\hspace*{-25mm}
\begin{minipage}[h]{.47\textwidth}
\includegraphics[height=9.5cm,width=9.5cm]{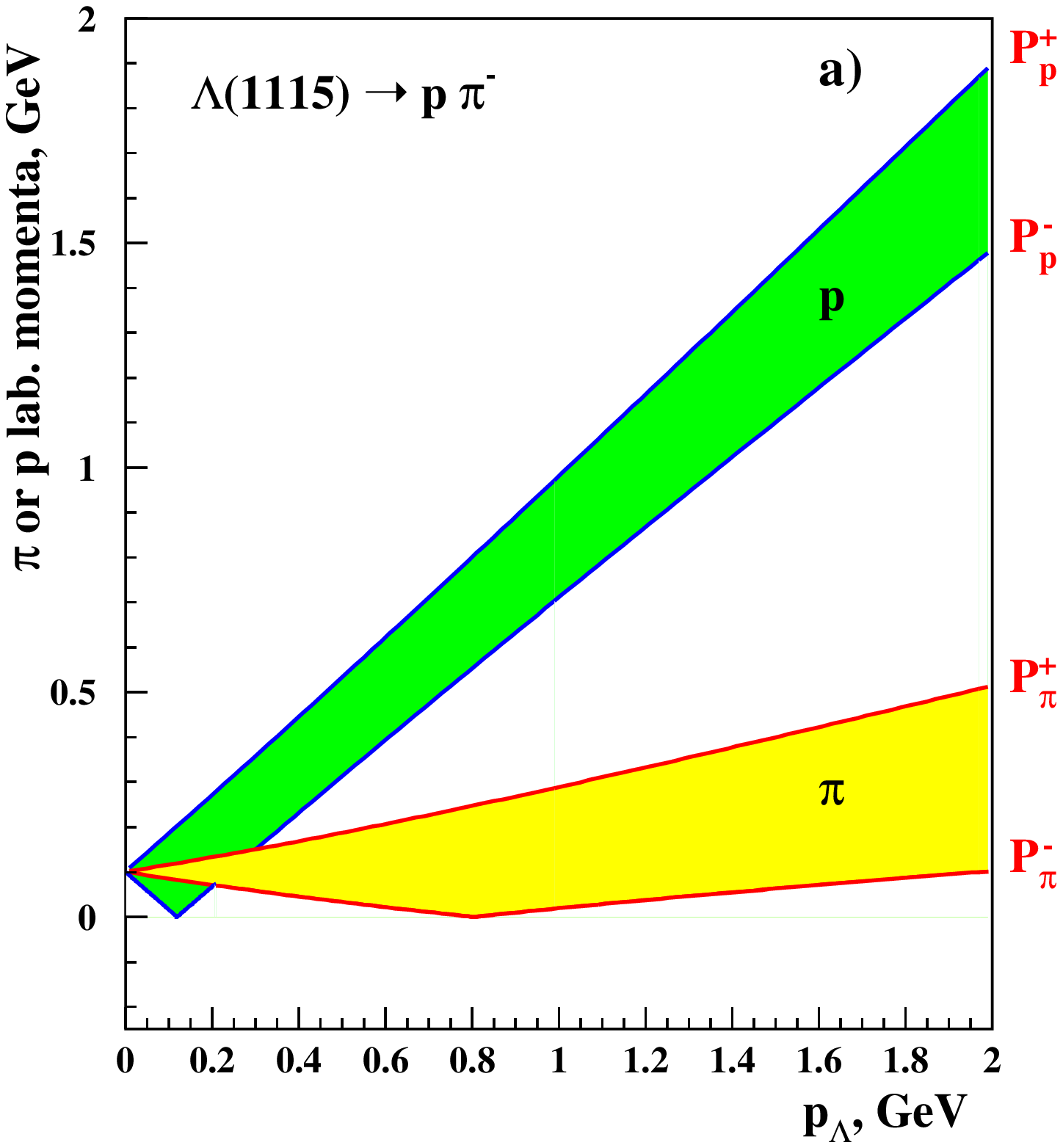}
\end{minipage}	\hspace*{+10mm}
\begin{minipage}[h]{.47\textwidth}
\includegraphics[height=9.5cm,width=9.5cm]{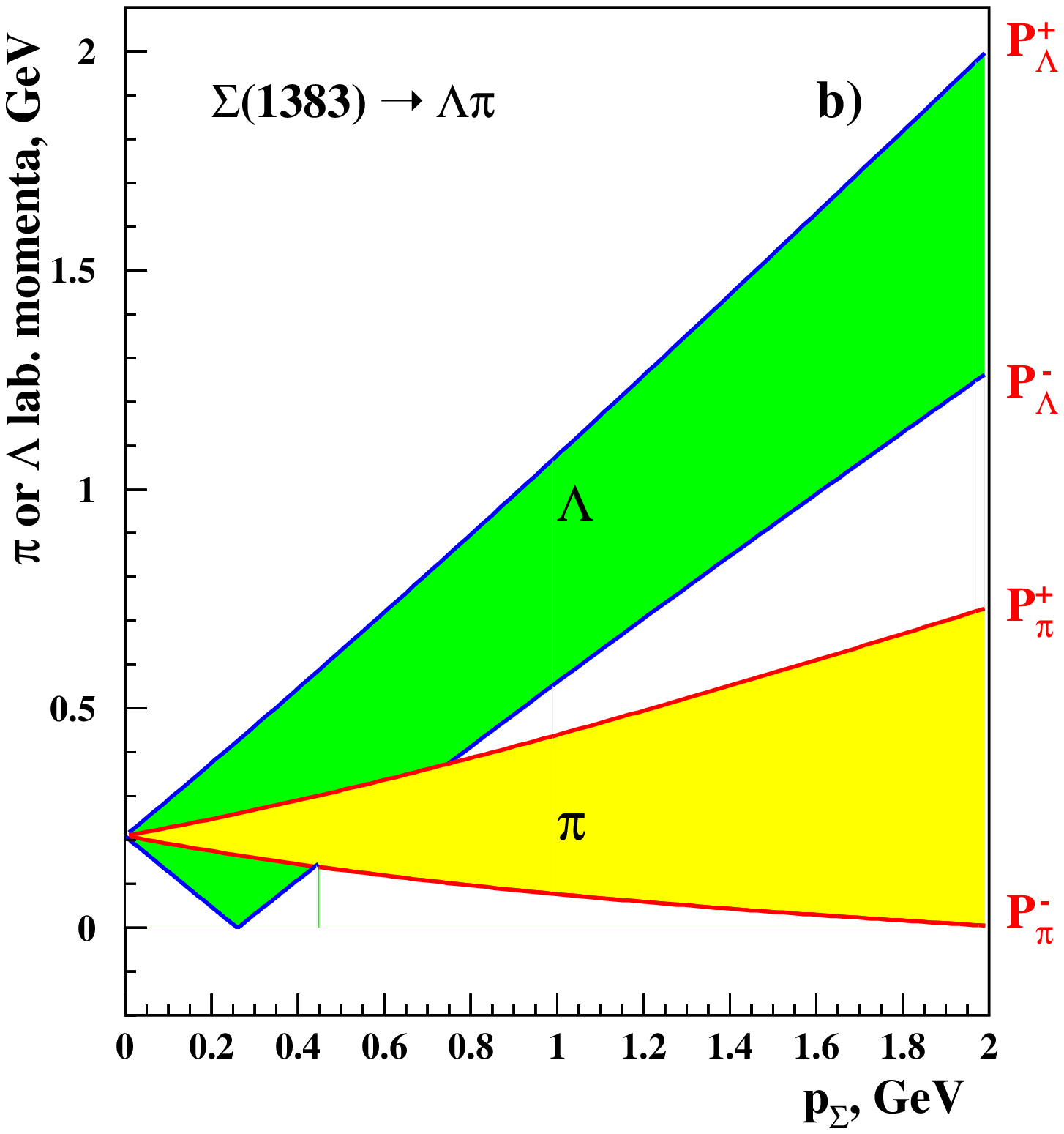}
\end{minipage}

\vspace*{-6mm}
\hspace*{-25mm}
\begin{minipage}[h]{.47\textwidth}
\includegraphics[height=9.5cm,width=9.5cm]{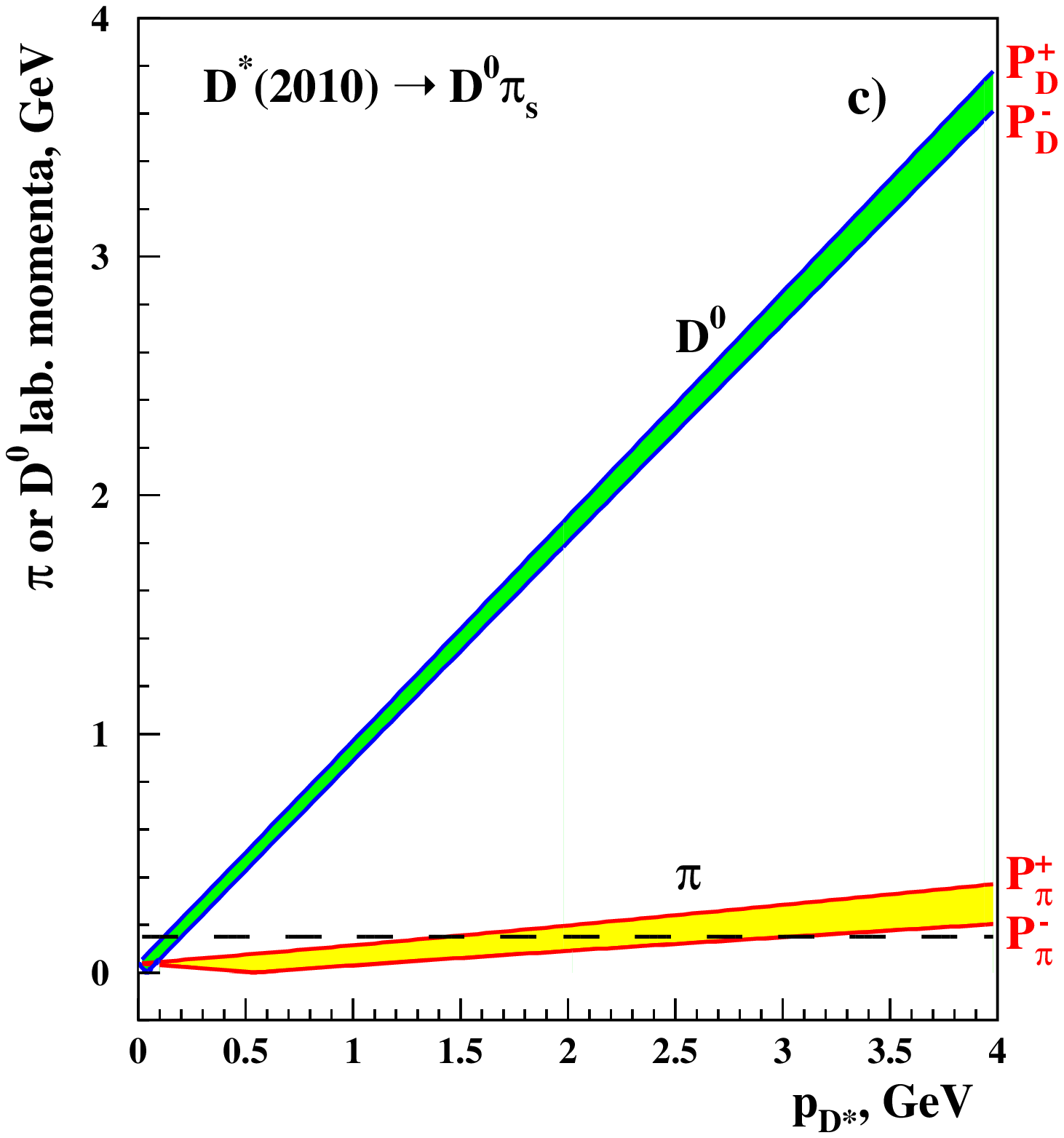}
\end{minipage}	\hspace*{+10mm}
\begin{minipage}[h]{.47\textwidth}
\includegraphics[height=9.5cm,width=9.5cm]{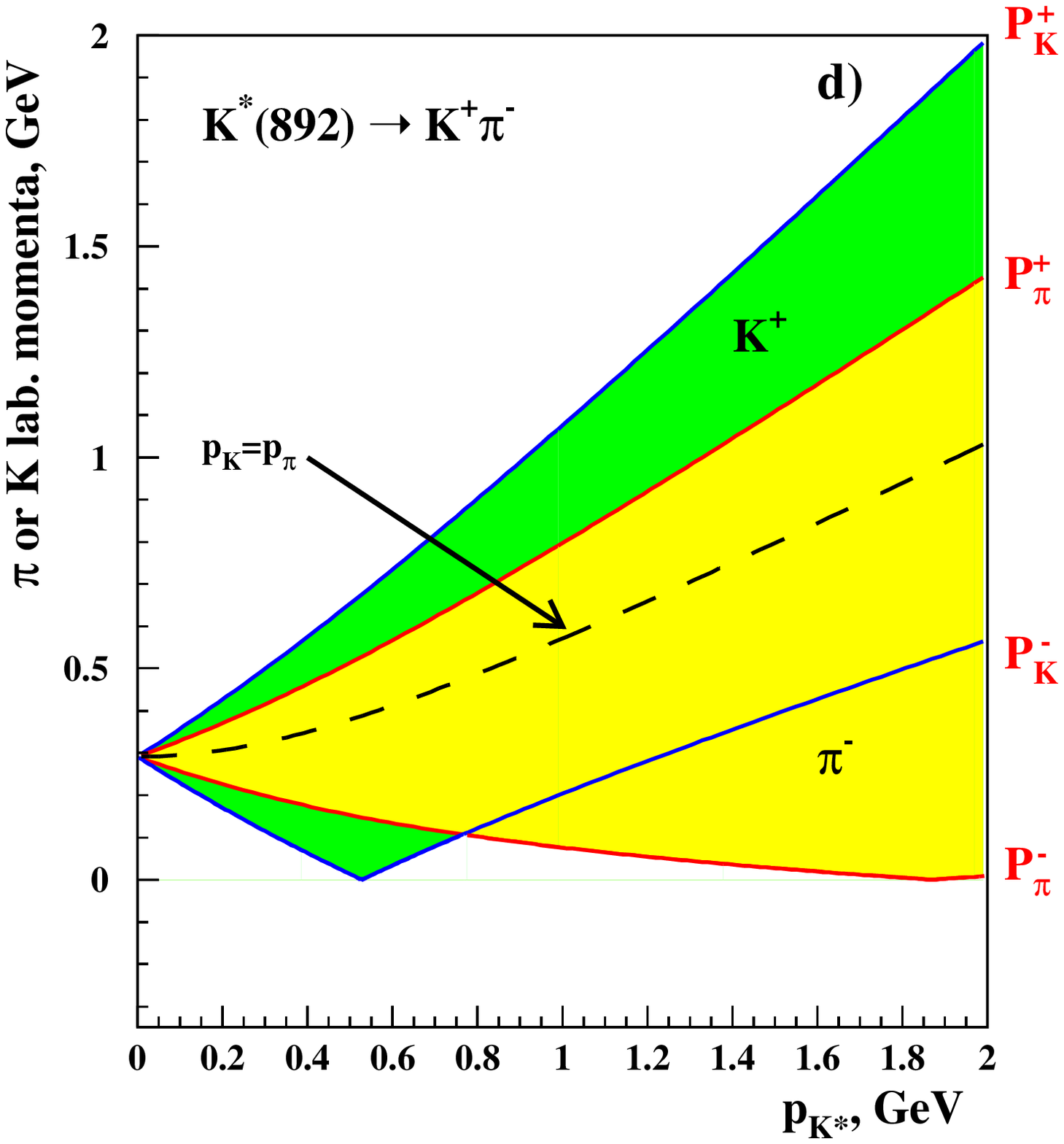}
\end{minipage}

%\begin{center}
\noindent
{\bf Figure~1:}{\it  Phase space $m$-bands of the particle {\bf a}
and {\bf b} in the decays $R\rightarrow {\bf a + b}$ as a function of the
resonance momentum $P_R$. The dashed line on (c)  corresponds to the momentum cut 
$p_{\pi}=0.15$ GeV (See Section 3). The dashed line on (d) is correspond 
to Eq.(\ref{eq:10}). }
%\end{center}

%\vspace*{10mm}

\newpage

\hspace*{-25mm}
\begin{minipage}[h]{.47\textwidth}
\includegraphics[height=9.5cm,width=9.5cm]{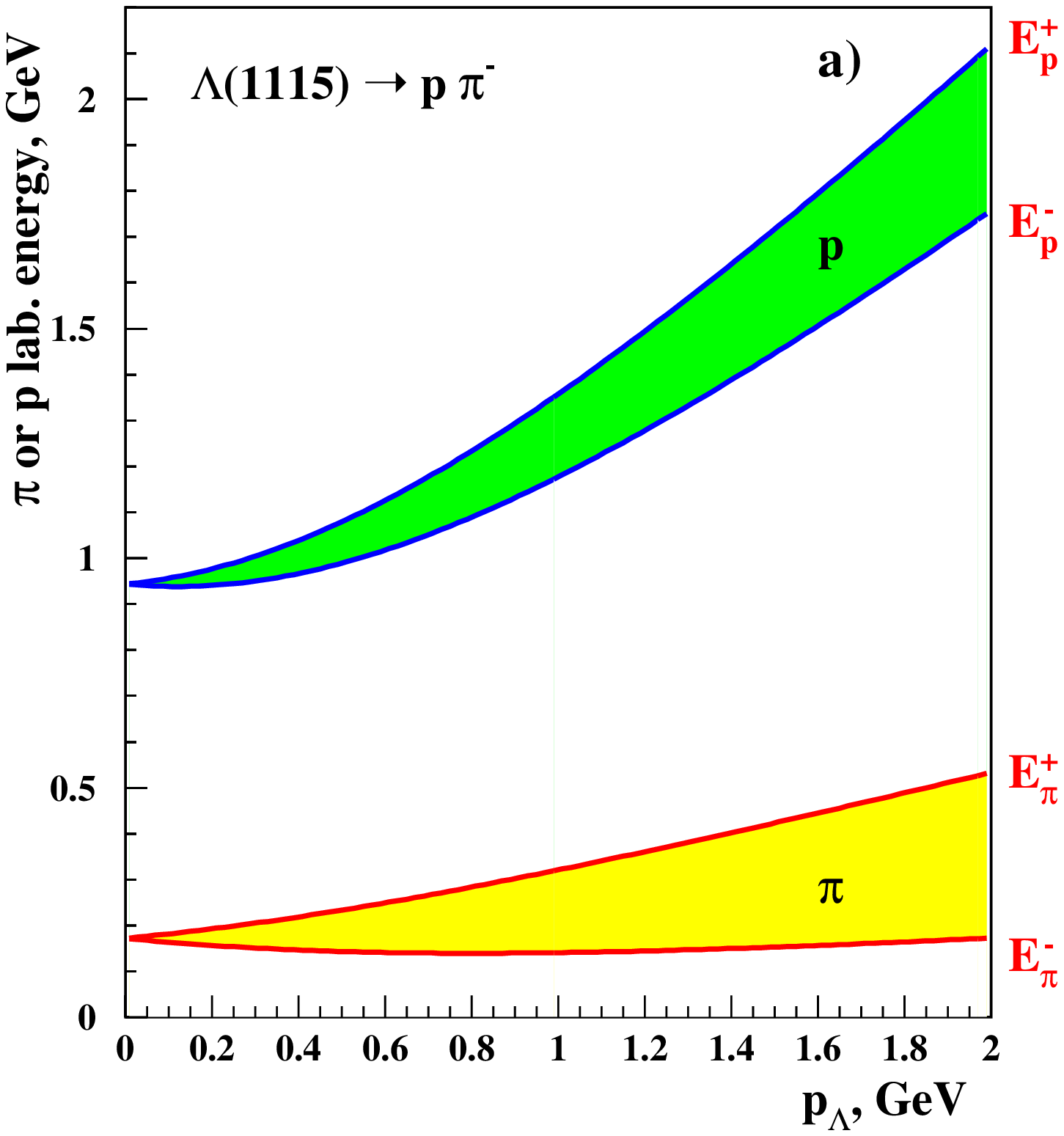}
\end{minipage}	\hspace*{+10mm}
\begin{minipage}[h]{.47\textwidth}
\includegraphics[height=9.5cm,width=9.5cm]{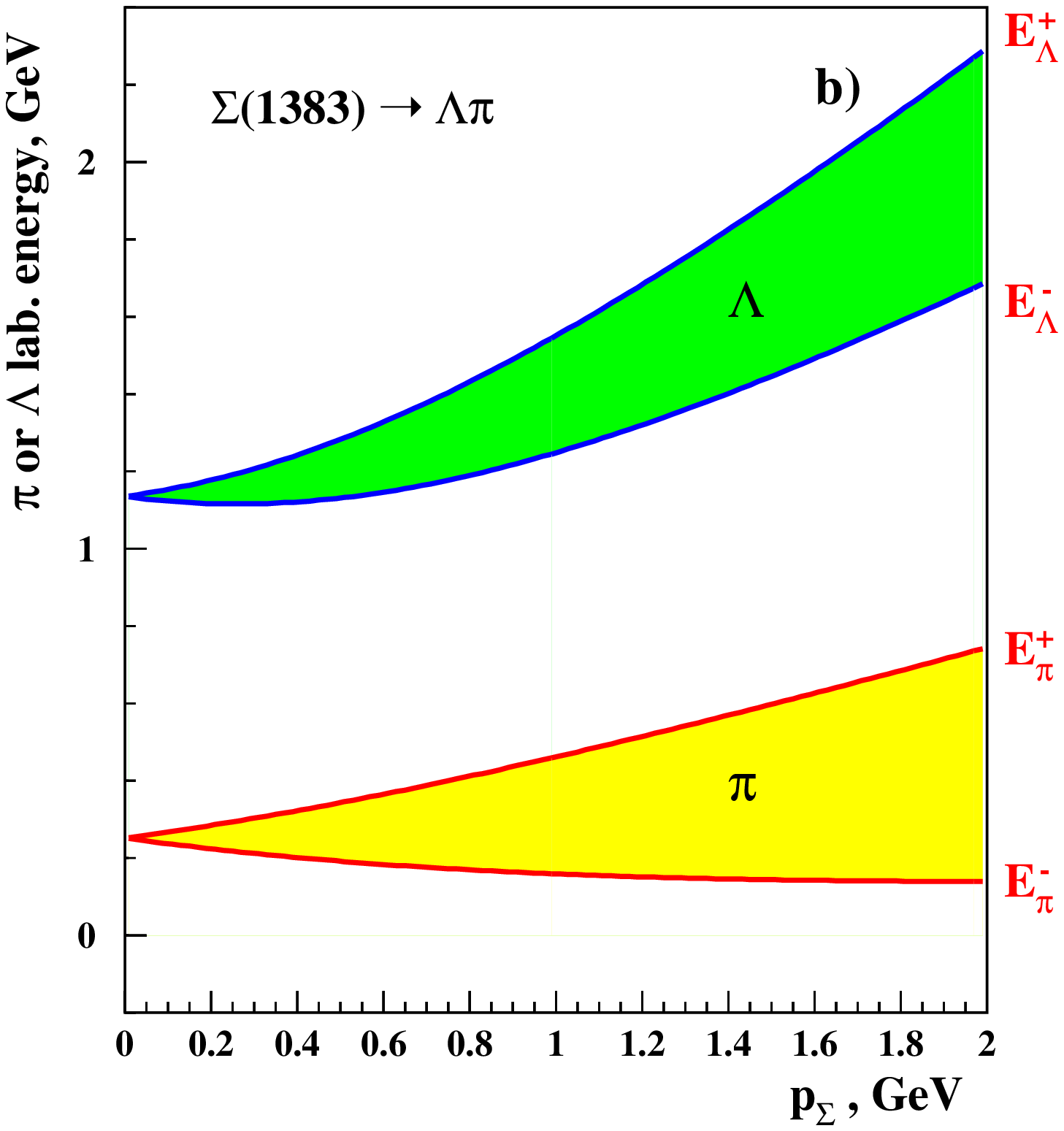}
\end{minipage}

\vspace*{-6mm}
\hspace*{-25mm}
\begin{minipage}[h]{.47\textwidth}
\includegraphics[height=9.5cm,width=9.5cm]{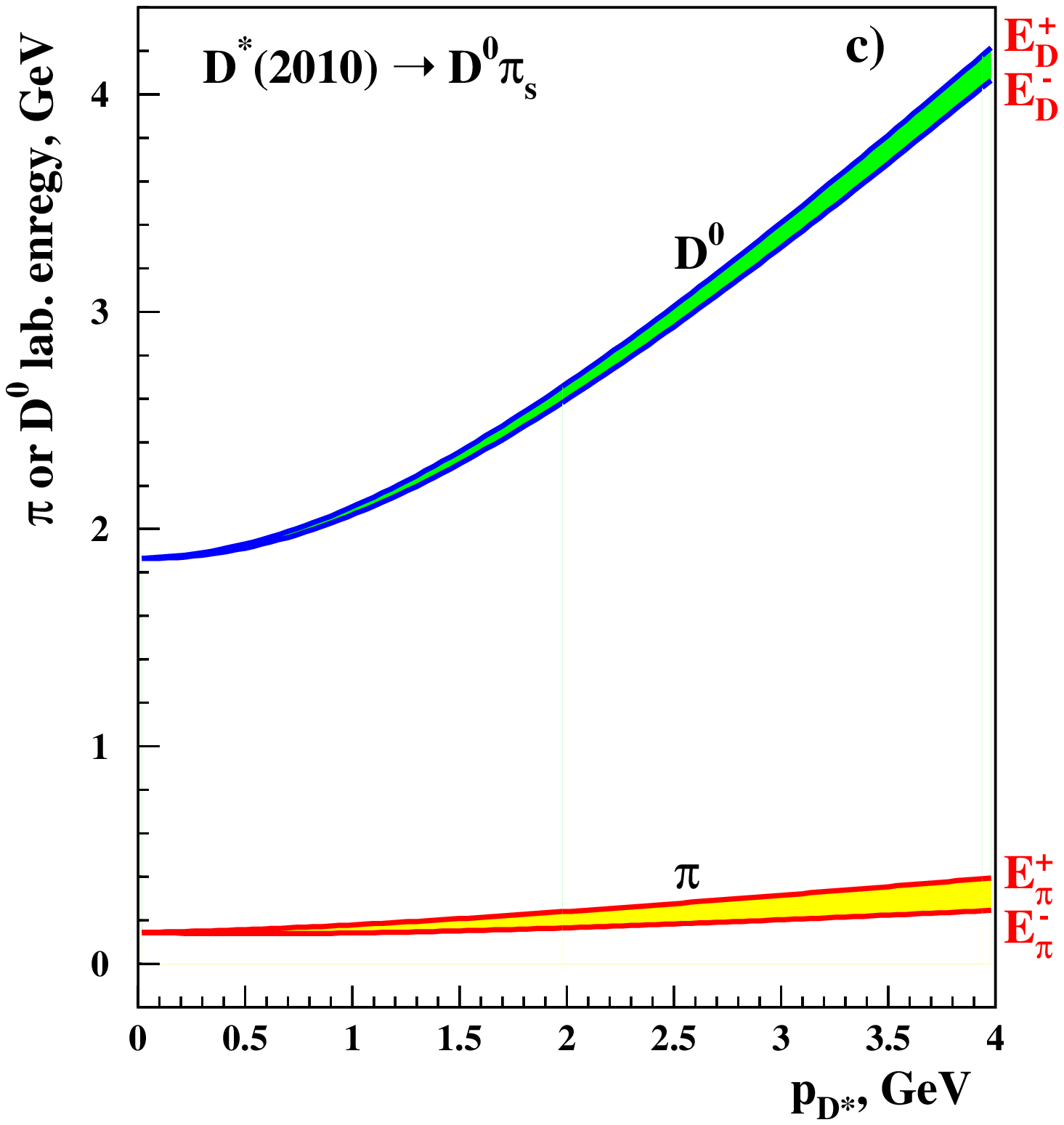}
\end{minipage}	\hspace*{+10mm}
\begin{minipage}[h]{.47\textwidth}
\includegraphics[height=9.5cm,width=9.5cm]{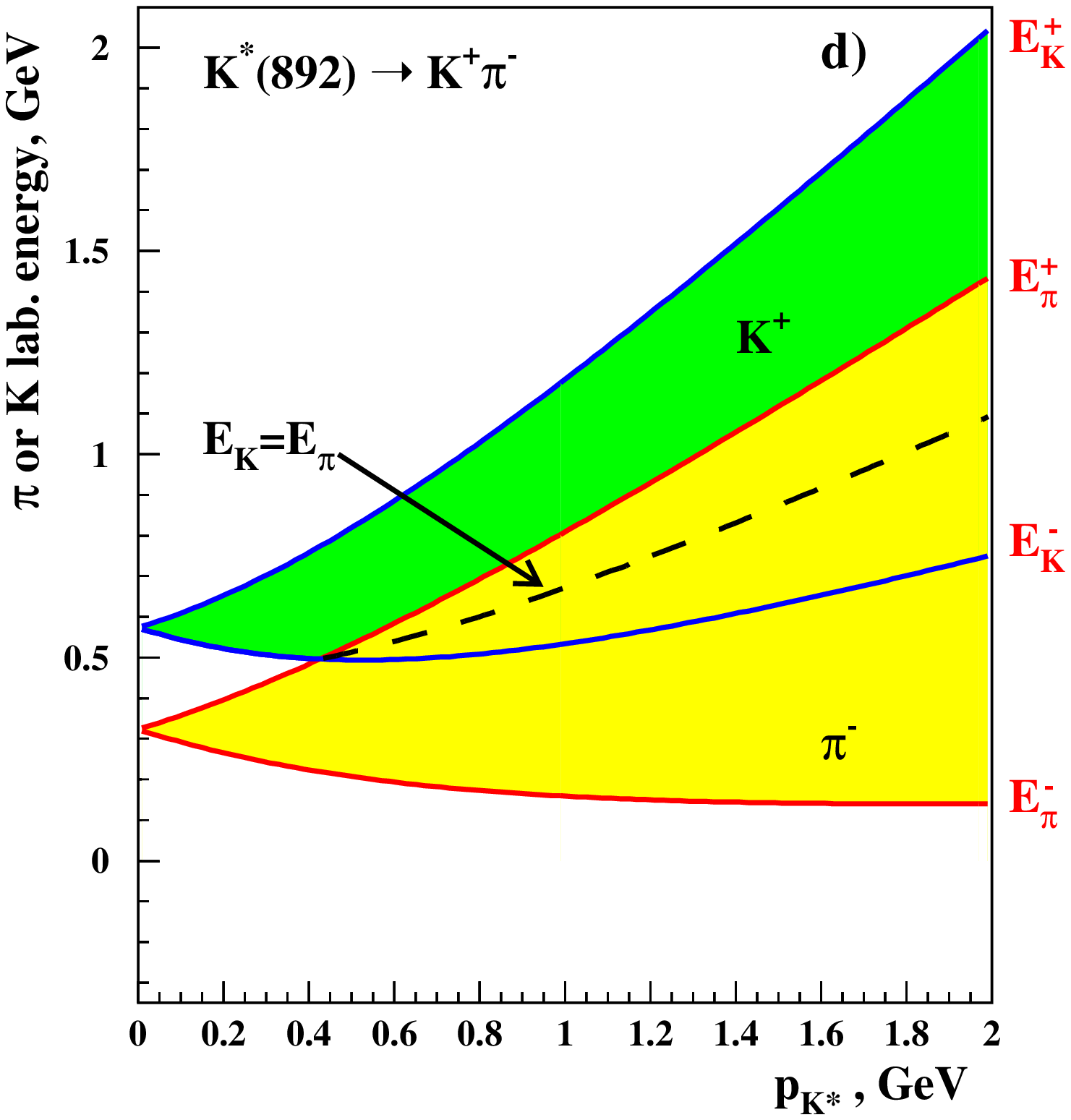}
\end{minipage}

%\begin{center}
\noindent
{\bf Figure~2:}{\it  Phase space $E$-bands in $R\rightarrow {\bf a + b}$ decays
as a function of the resonance momentum $P_R$. The dashed line on (d)  corresponds 
to Eq.(\ref{eq:18}). }
%\end{center}

\newpage

\vspace*{-15mm}
\hspace*{-25mm}
\begin{minipage}[h]{.47\textwidth}
\includegraphics[height=8.cm,width=9.cm]{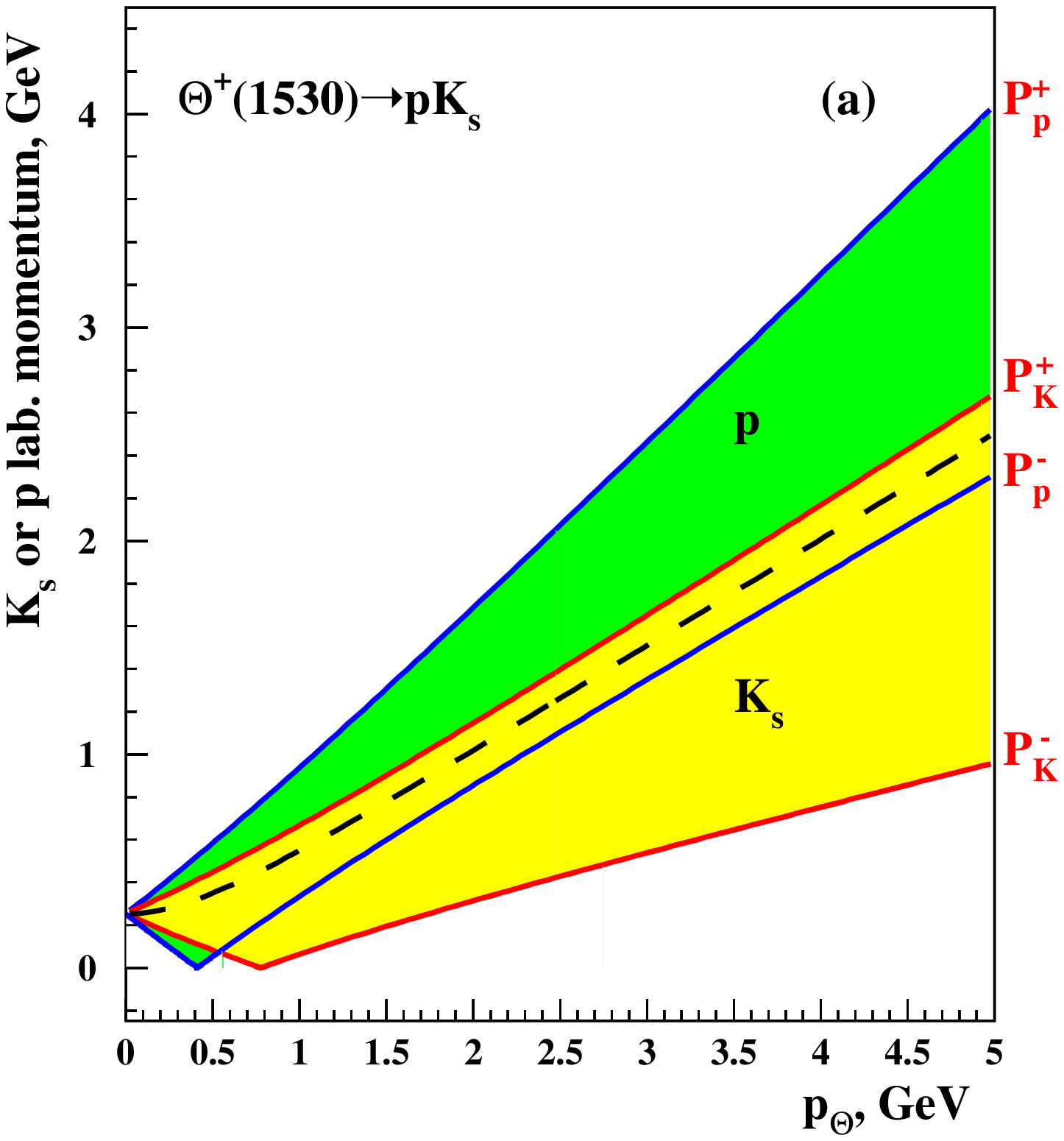}
\end{minipage}	\hspace*{+10mm}
\begin{minipage}[h]{.47\textwidth}
\includegraphics[height=8.cm,width=9.cm]{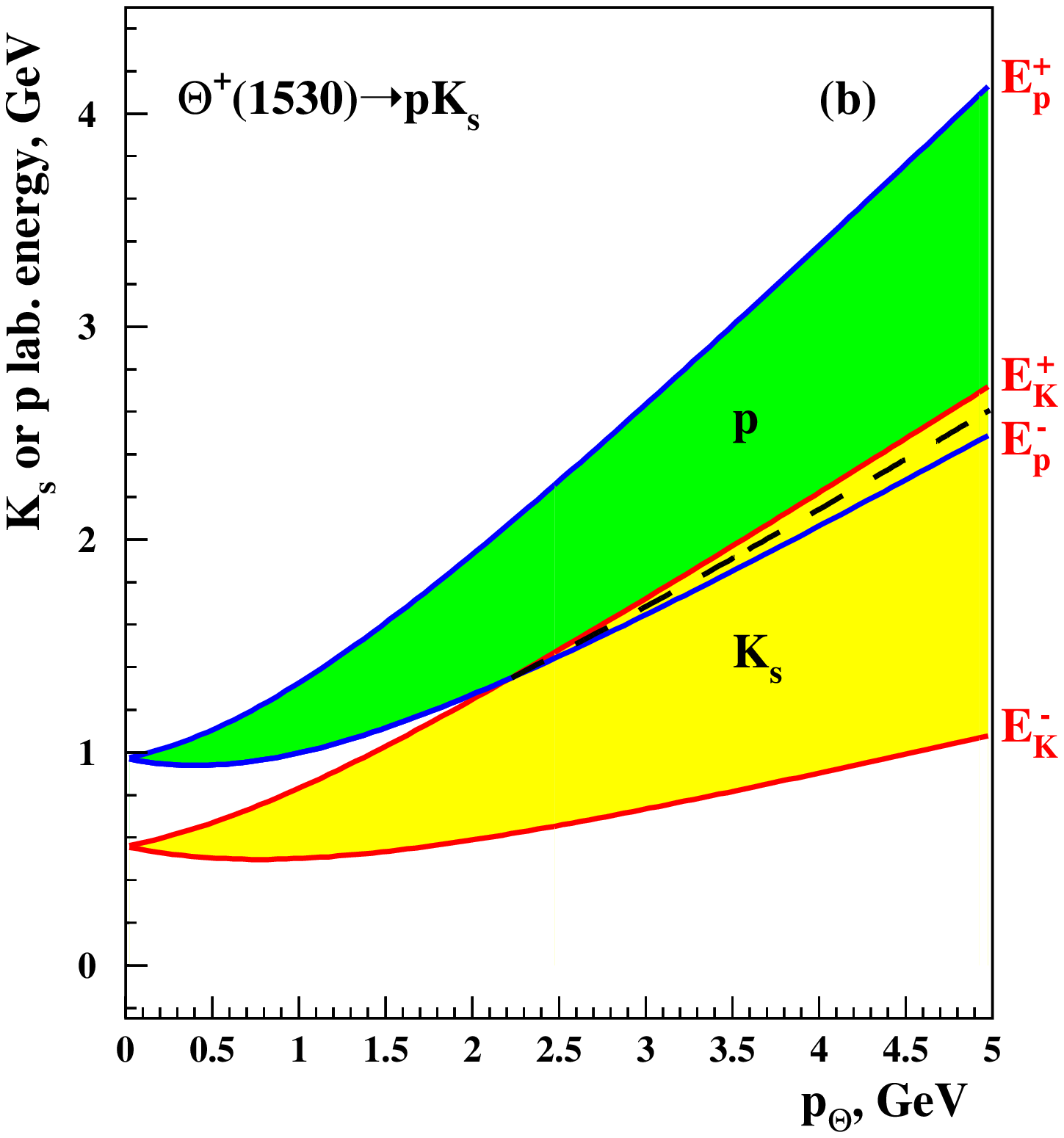}
\end{minipage}

\vspace*{-8mm}
\hspace*{-25mm}
\begin{minipage}[h]{.47\textwidth}
\includegraphics[height=8.cm,width=9.cm]{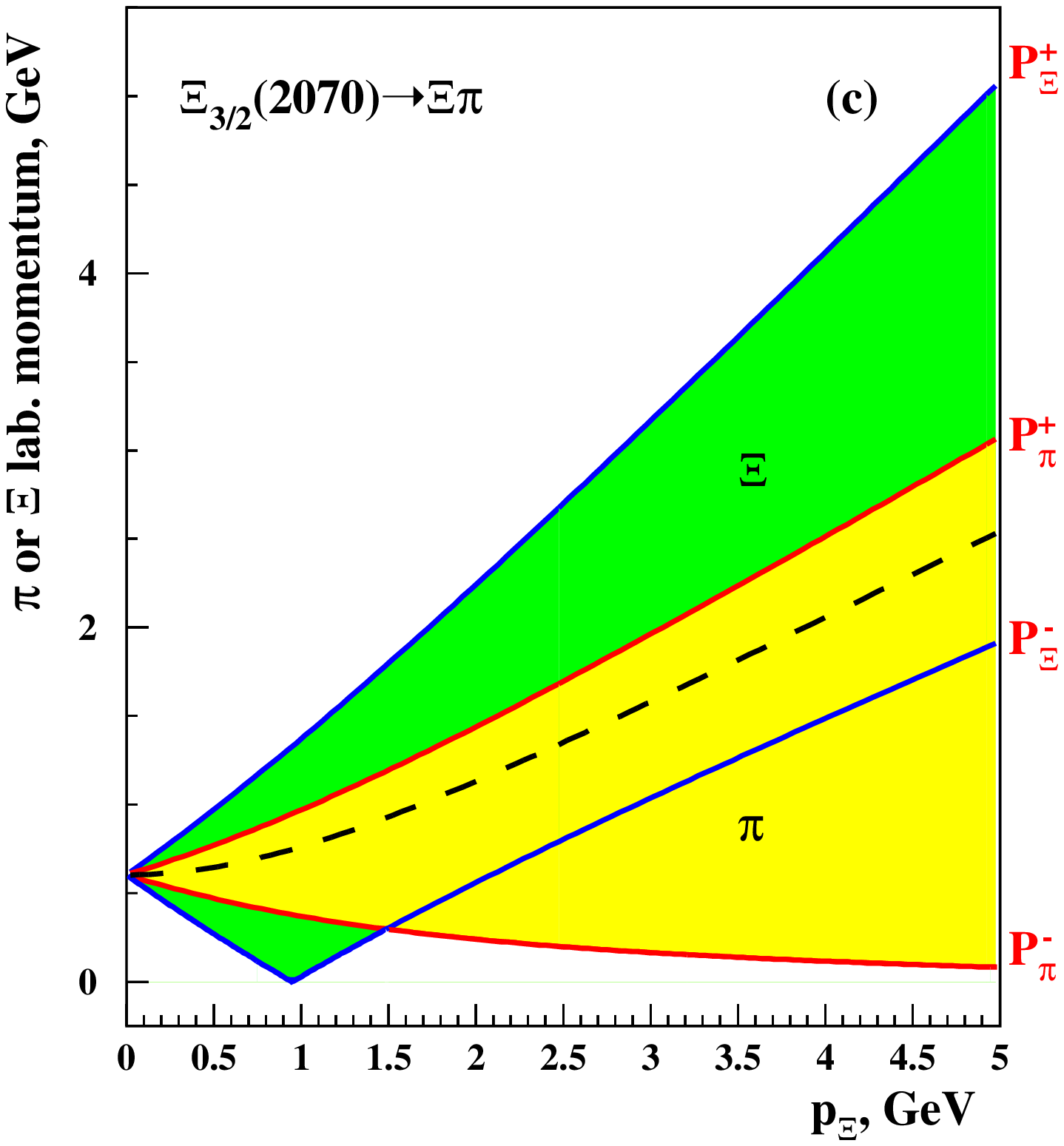}
\end{minipage}	\hspace*{+10mm}
\begin{minipage}[h]{.47\textwidth}
\includegraphics[height=8.cm,width=9.cm]{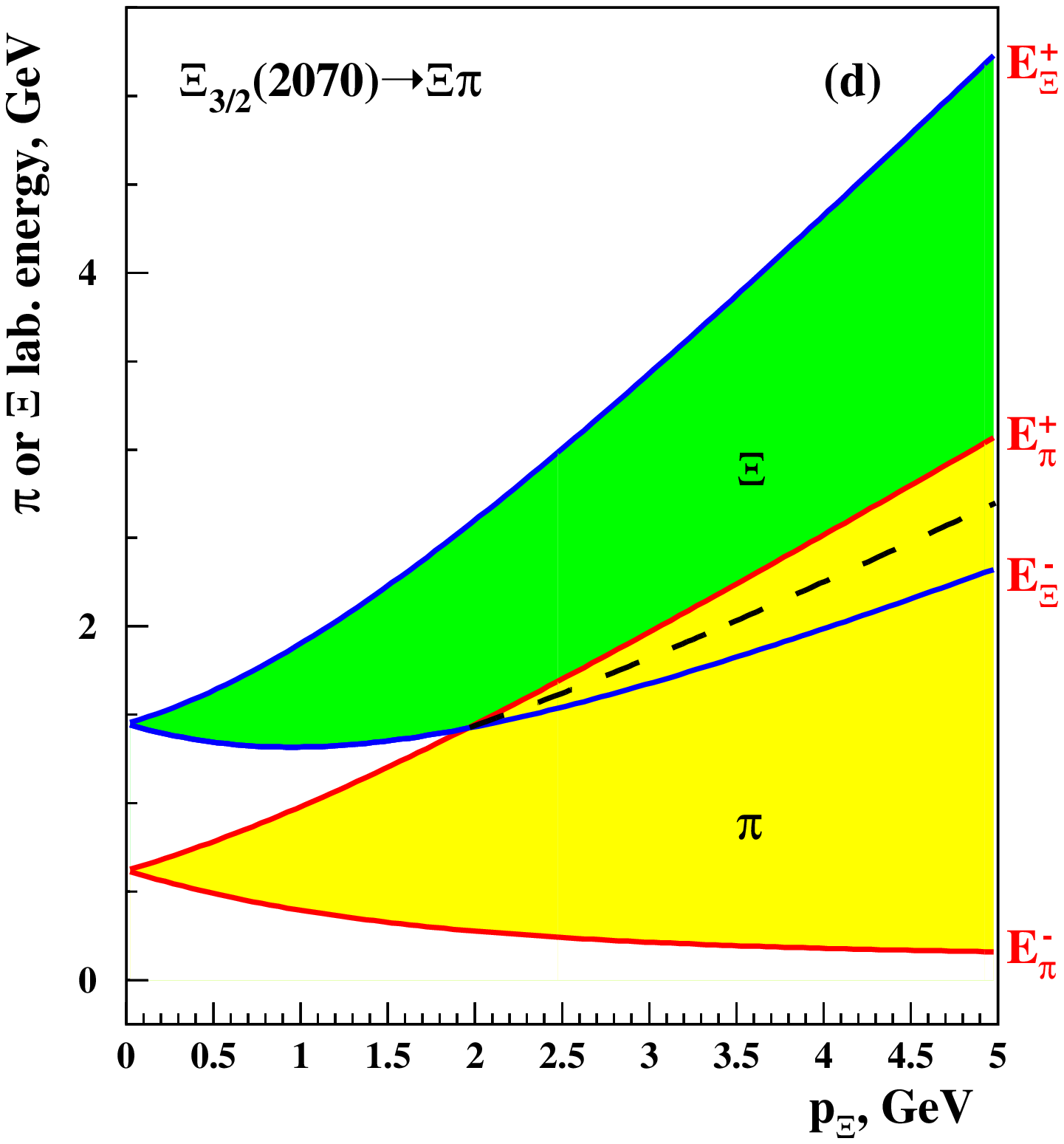}
\end{minipage}

\vspace*{-8mm}
\hspace*{-25mm}
\begin{minipage}[h]{.47\textwidth}
\includegraphics[height=8.cm,width=9.cm]{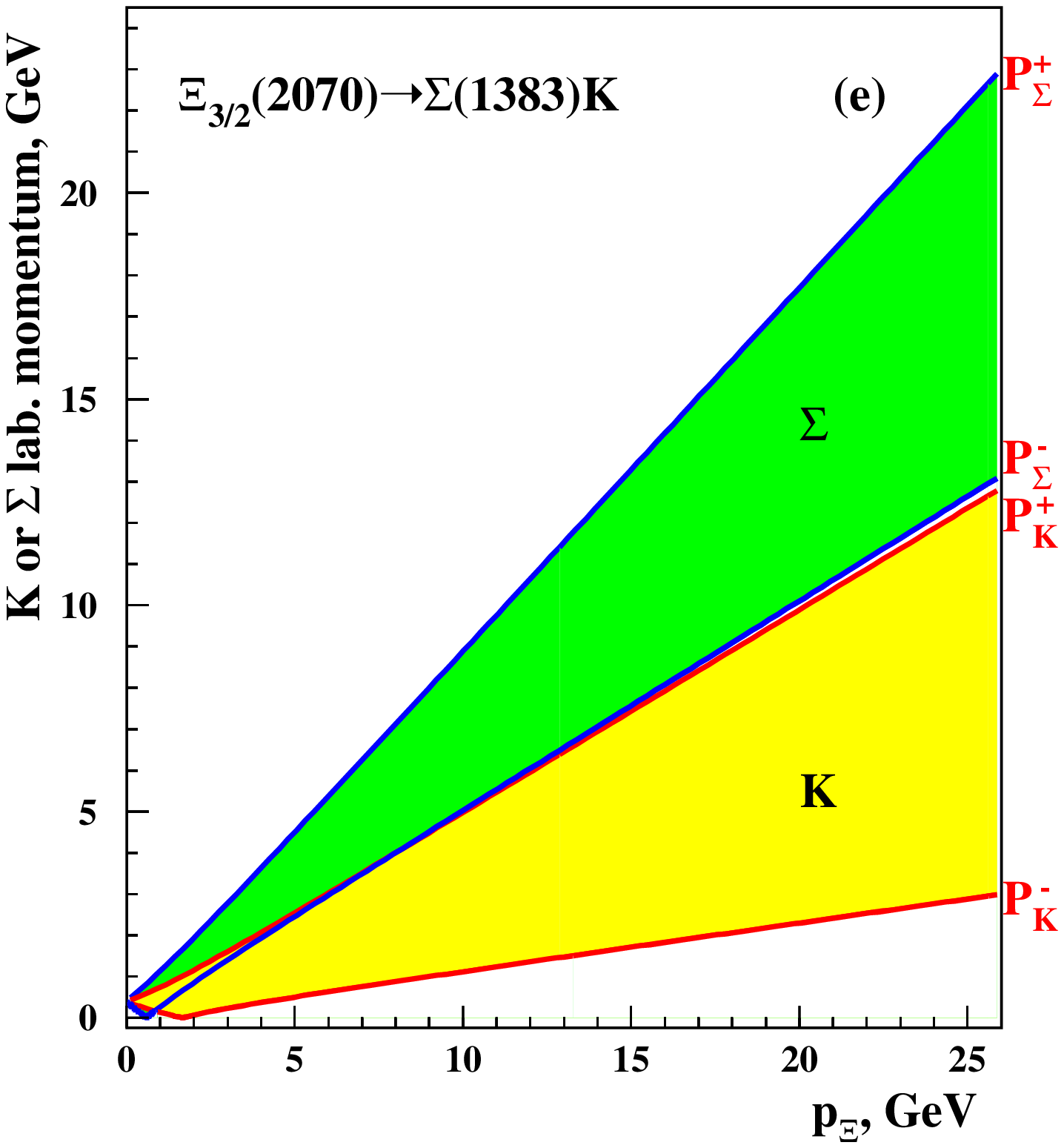}
\end{minipage}	\hspace*{+10mm}
\begin{minipage}[h]{.47\textwidth}
\includegraphics[height=8.cm,width=9.cm]{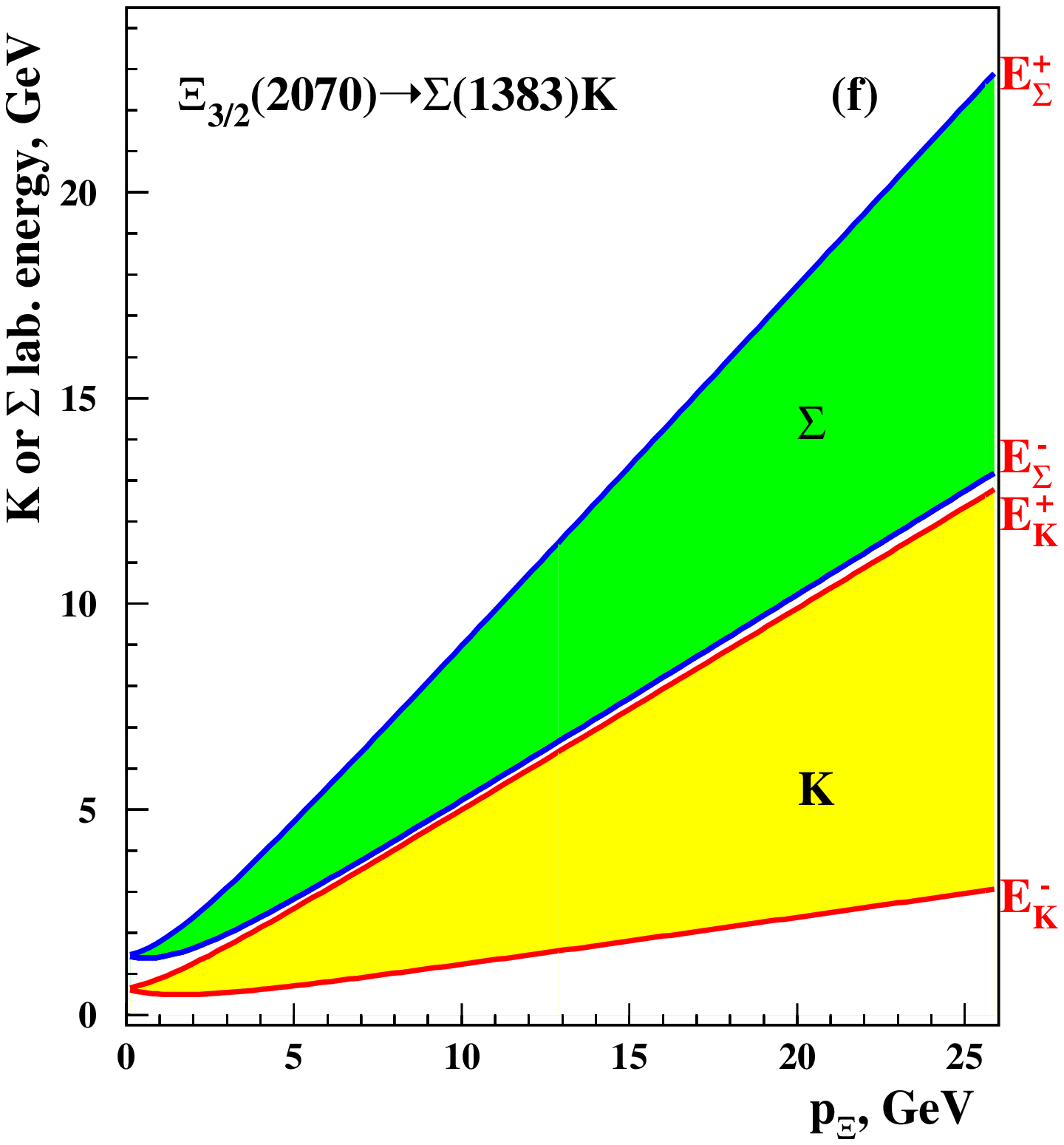}
\end{minipage}

\noindent
{\bf Figure~3:}{\it Phase space $m$-bands  ((a), (c), (e)) and $E$-bands ((b), (d), (f)) 
in decays of the pentaquark states $\Theta(1530)$ and $\Xi_{3/2}(2070)$.
The dashed lines  correspond to Eqs.(\ref{eq:10}) and (\ref{eq:18}).}

\newpage
\section{Remark about $D^*$ reconstruction}
Charged particles are tracked in a tracking detector (TD). The resolution
of the transverse momentum of a track  traversing the TD
is parametrized  by $\sigma(p_T)/p_T\,=\,A p_T\bigoplus B\bigoplus C/p_T$, 
with $p_T$ being the track transverse momentum (in GeV). 
The coefficients $A,\,B$ and $C$   characterize the resolution of the TD. 
Usually, to increase the momentum resolution,  only tracks 
with $p_T>0.12-0.15$ GeV  are selected. That cut, as shown in Fig. 1c, make 
impossible the reconstruction of $D^*$ mesons with momenta lower than 1.8-2.0 GeV.
The  $m$-band of $\pi$ mesons is very narrow and grows rather slowly with $P_R$.
Tracks with momenta greater than $p^+_{\pi}(P_{D^*})$ belong to the background.
This property can be used to suppress the background contribution to the 
distribution of the mass difference,  $\Delta M=M(K\pi\pi_s )-M(K\pi )$, 
%for $D^*$ candidates 
by applying to the momentum of the soft pion ($\pi_s$) the following  cut
\beq 
p_{\pi_s}< p_0 + p^+_{\pi}(P_{D^*})
\label{eq:20}
\eeq
where  $p_0$=0.0-0.3 GeV is a some  shift from the pion  $m$-band.
In the decay $D^*\rightarrow D^0\pi$,  the rest frame momentum is small, $P^*=$0.038 GeV,
and the  $D^*$ momentum can be estimated with (\ref{eq:4})
by means of the reconstructed $D^0$ momentum, 
\beq 
P_{D^*}=\frac{M_{D^*}}{M_{D^0}}\cdot P_{D^0}
\label{eq:21}
\eeq
Thus, from Eqs.(\ref{eq:21}) and  (\ref{eq:4}) one get for (\ref{eq:20})
\beq 
p_{\pi_s}< p_0 + \frac{(m_{\pi}+P^*)}{M_{D^0}}\cdot P_{D^0}
\label{eq:22}
\eeq
Instead of  (\ref{eq:22}) it is possible to apply another,  less strict cut
\beq 
p_{\pi_s}<  p^+_{\pi}(P_{max})\simeq \frac{(m_{\pi}+P^*)}{M_{D^*}}\cdot P_{max}
\label{eq:23}
\eeq
here $P_{max}$ is the right-hand edge  of the kinematic range of the $D^*$ candidates,
$P_{D^*}<P_{max} $.

\section{Pentaquark states}

Now we apply the results of the previous sections to new resonances
predicted \cite{DPP}  by Diakonov, Petrov and Polyakov  in the framework of the
chiral soliton model and detected 
both in the formation type \cite{5qfte1}  and in the production type
\cite{5qpte1} experiments.

\subsection{$\Theta(1530)\rightarrow N(939)+K(498)$}

With the mass value predicted for $\Theta^+$, $M_{\Theta}$=1.530 GeV,
and the masses\footnote{In calculations were used  mass values averaged over the isomultiplet.}
of decay products, $m_N$=0.939 GeV and $m_K$=0.498 GeV, the condition (\ref{eq:7})
is not satisfied. That implies the overlap of the $m$-bands at all $P_{\Theta}$ (Fig. 3a), 
as well as the overlap of the $E$-bands at $P_{\Theta}>\hat{P}_{\Theta}$=2.15 GeV (Fig. 3b).
The overlapping is not strong and both selectors (\ref{eq:9}) and  (\ref{eq:16})   work
with high efficiency (see Fig.~4). The $m$- and $E$-selectors 
were already successfully applied in searches for the $\Theta^+$ in   production-type 
\cite{5qpte1} and  formation-type \cite{5qfte2} experiments.

%\newpage
\hspace*{-25mm}
\begin{minipage}[h]{.47\textwidth}
\hspace*{+5mm}\includegraphics[height=9.5cm,width=9.5cm]{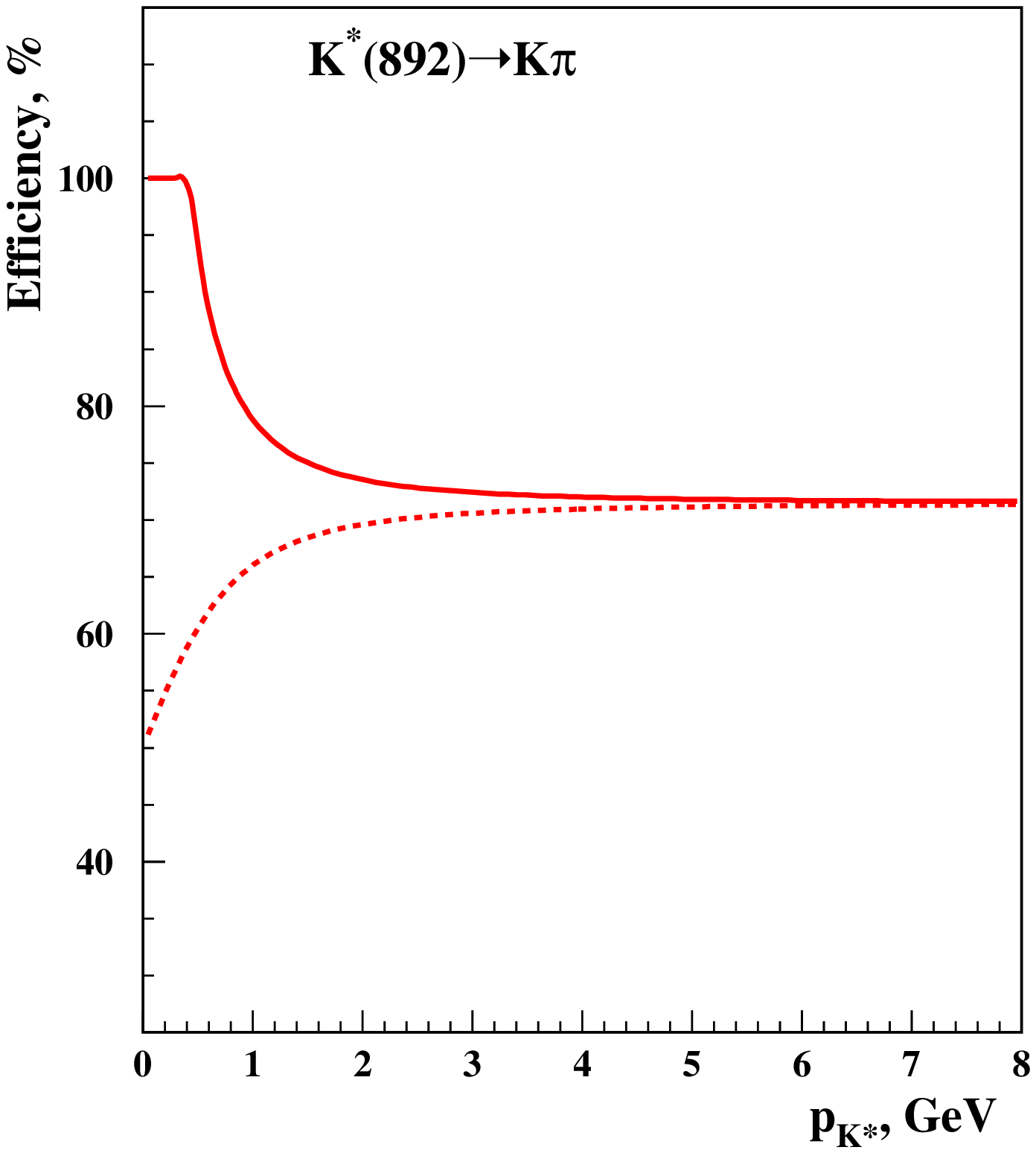}
\end{minipage}	\hspace*{+10mm}
\begin{minipage}[h]{.47\textwidth}
\includegraphics[height=9.5cm,width=9.5cm]{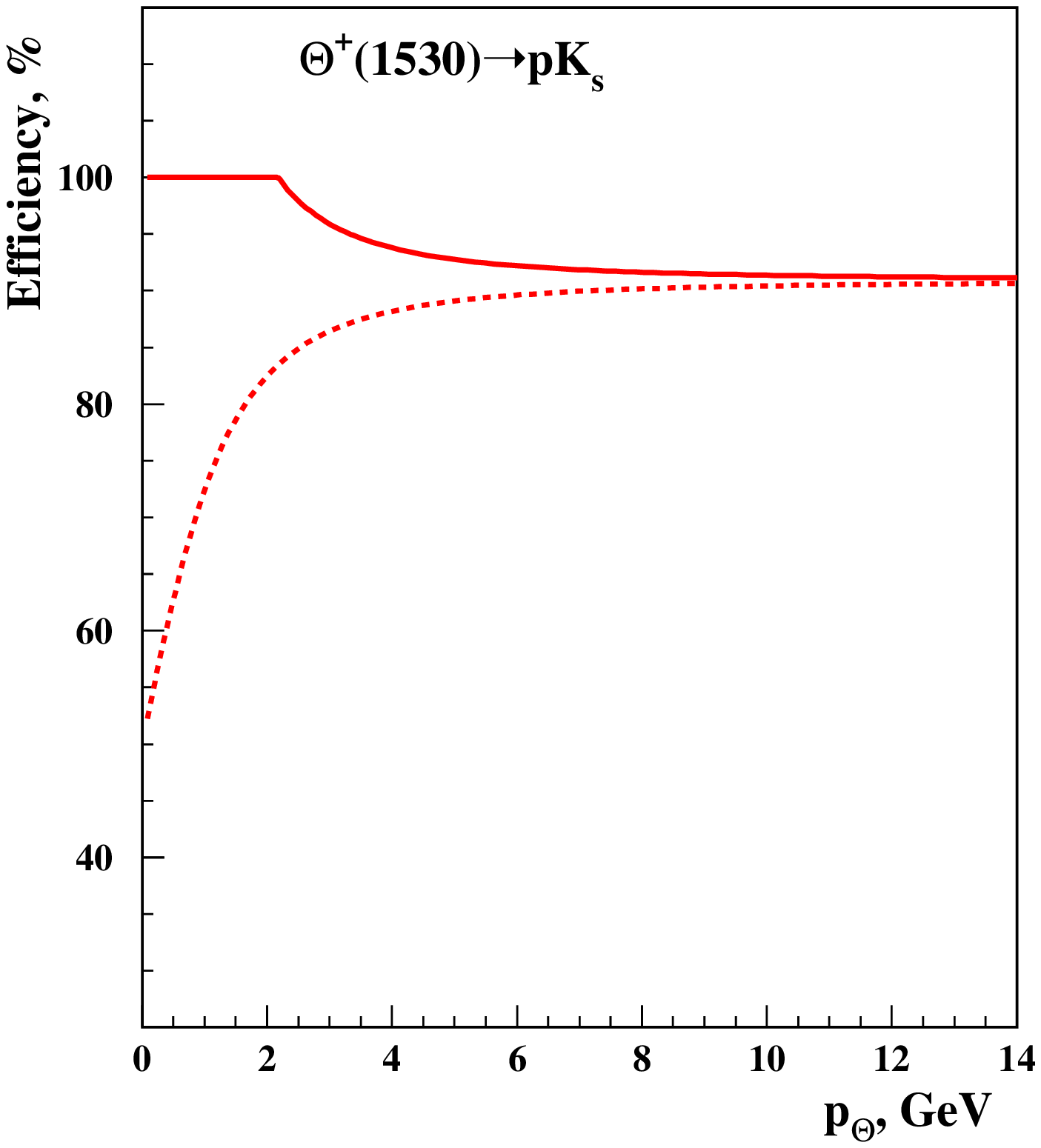}
\end{minipage}

\vspace*{-6mm}
\hspace*{-25mm}
\begin{minipage}[h]{.47\textwidth}
\hspace*{+5mm}\includegraphics[height=9.5cm,width=9.5cm]{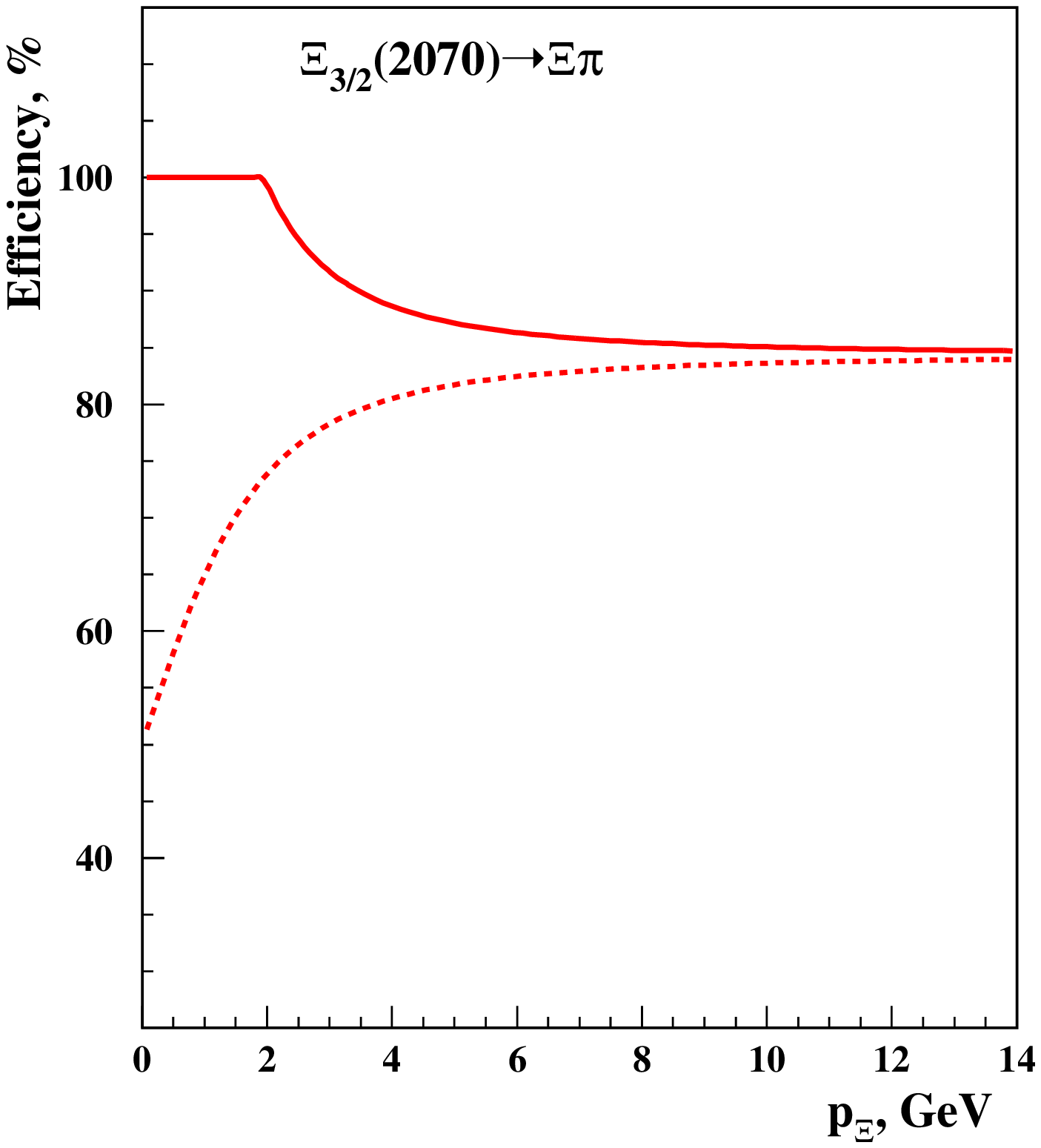}
\end{minipage}	\hspace*{+10mm}
\begin{minipage}[h]{.47\textwidth}
\includegraphics[height=9.5cm,width=9.5cm]{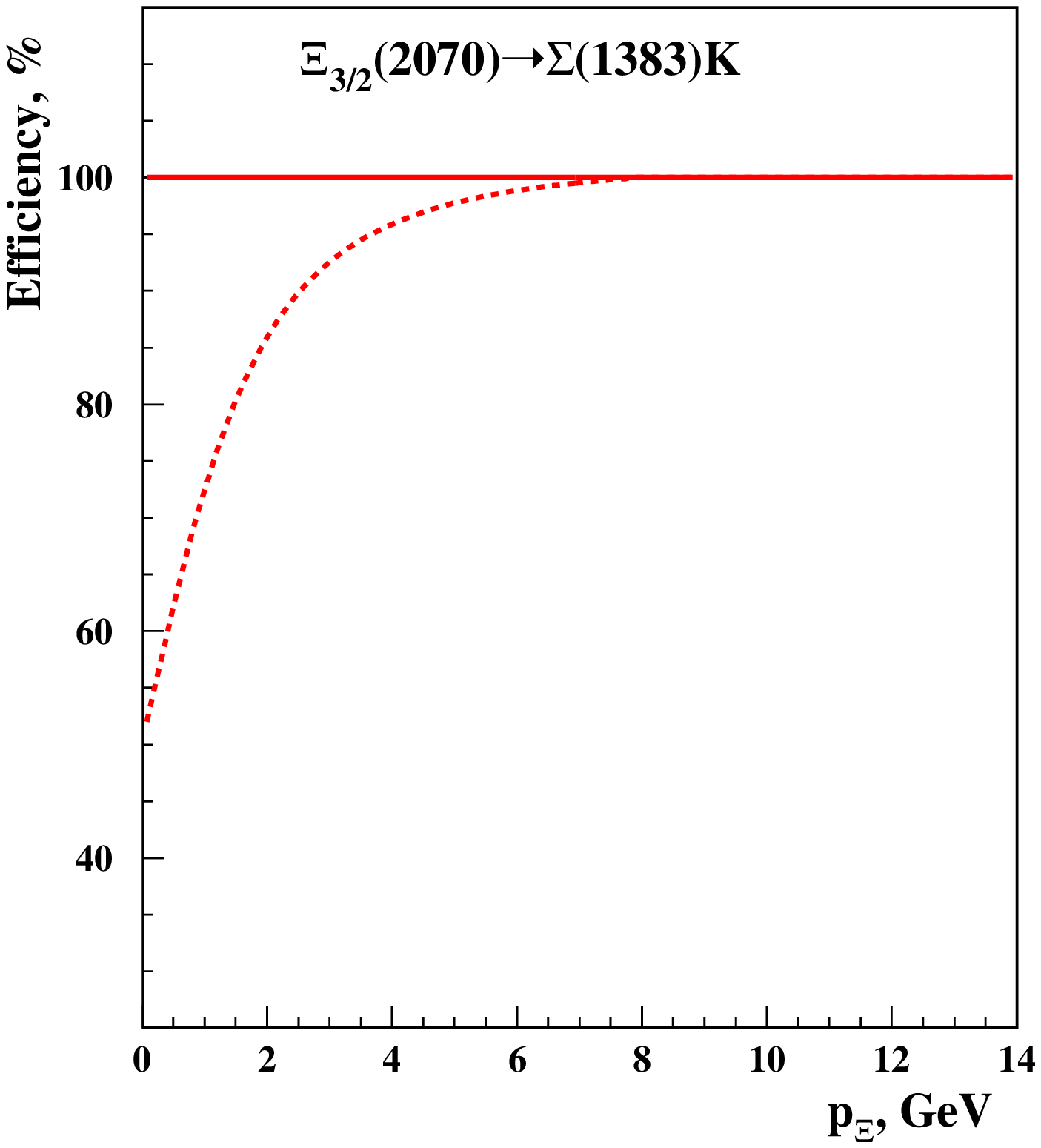}
\end{minipage}

\noindent
{\bf Figure~4:}{\it Efficiency of the selection rule $p_{\bf a}>p_{\bf b}$ 
(dashed line) and $E_{\bf a}>E_{\bf b}$ (full line) at different $P_R$
in decays of $K^*(892)$, $\Theta(1530)$ and $\Xi_{3/2}(2070)$ resonances.
}

\vspace*{+10mm}

\subsection{$\Xi_{3/2}(2070)\rightarrow \Xi (1318) + \pi(139)$}
A baryonic state $\Xi$ is a weakly decaying particle. In a cascade decay,
the vertex of the decay $\Xi\rightarrow \Lambda \pi$ is separated from the primary
vertex, similarly to $\Lambda\rightarrow p\pi^-$ decays. This property is

\hspace*{-25mm}
\begin{minipage}[h]{.47\textwidth}
\hspace*{+5mm}\includegraphics[height=9.5cm,width=9.5cm]{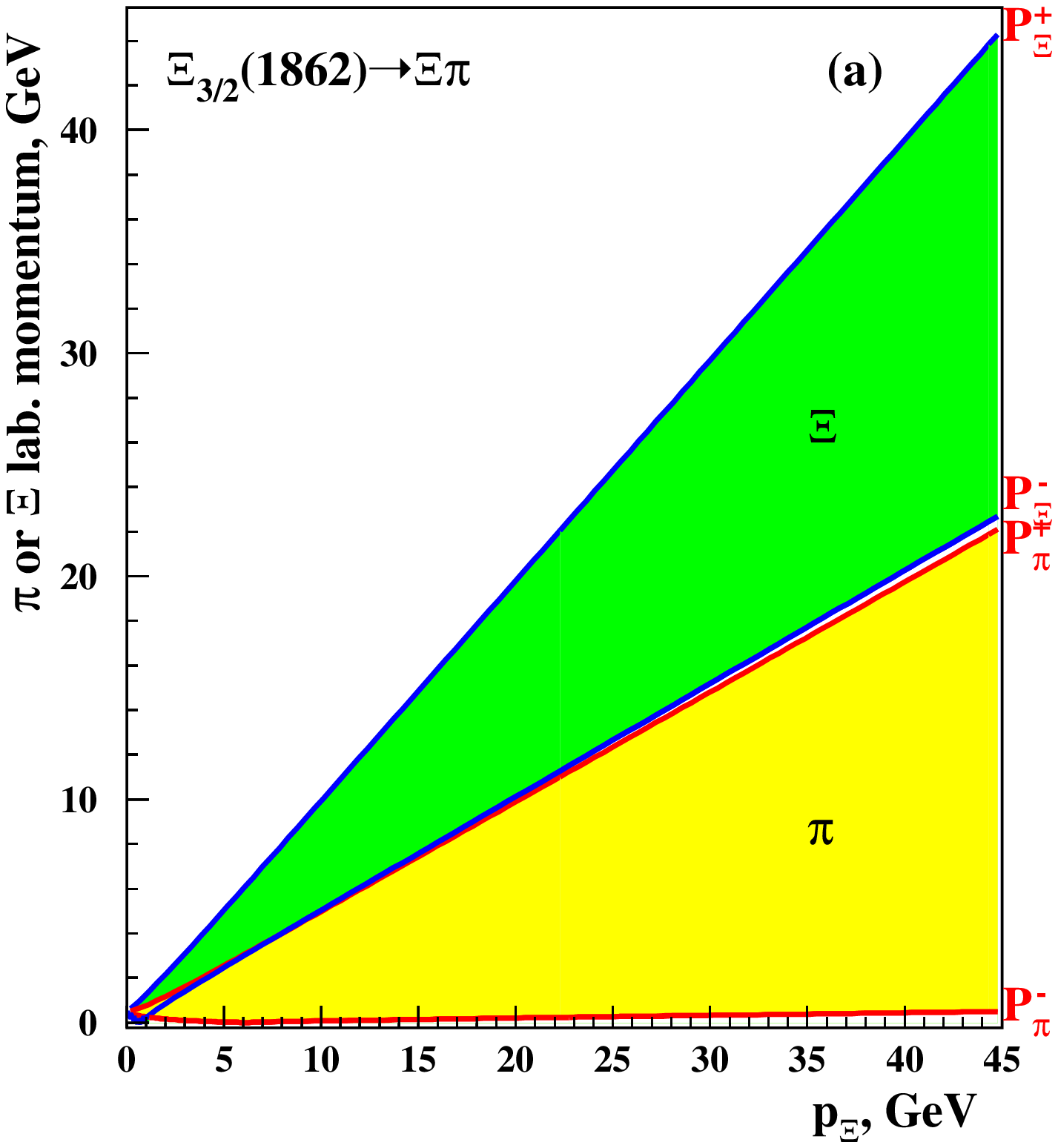}
\end{minipage}	\hspace*{+10mm}
\begin{minipage}[h]{.47\textwidth}
\includegraphics[height=9.5cm,width=9.5cm]{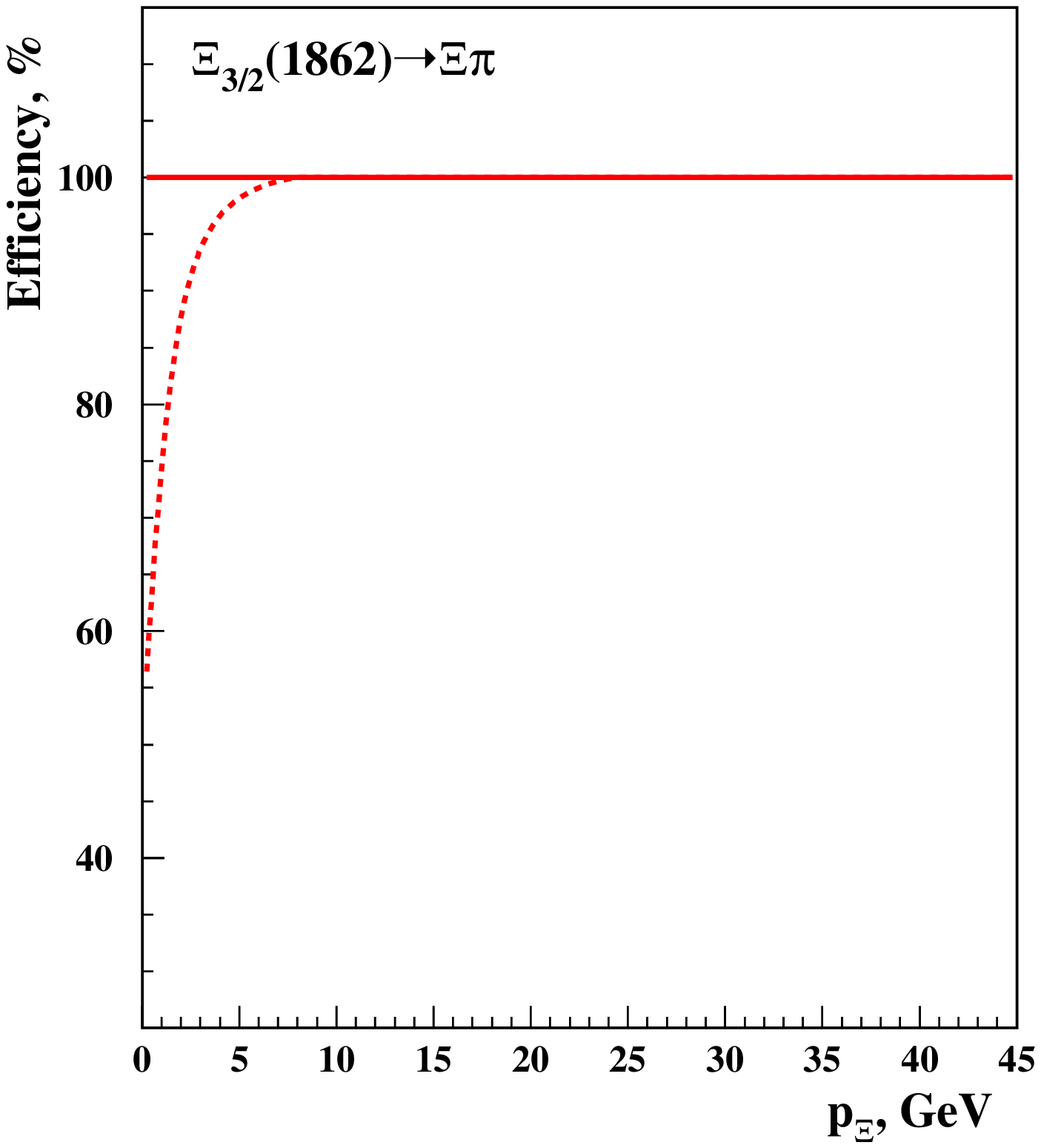}
\end{minipage}

\noindent
{\bf Figure~5:}{\it The $m$-bands and the  efficiency of the $m$- and 
$E$-selectors for the  $\Xi_{3/2}\rightarrow \Xi+\pi$ decay with the $\Xi_{3/2}$ mass reported by the NA49 collaboration
\cite{NA49}.
}

\vspace*{+7mm}
\noindent
used to suppress combinations which do not
result from a $\Xi$ decay and helps to reconstruct the  $\Xi$ candidate and
its  invariant mass.

In spite of the large mass asymmetry, $m_{\Xi}\gg m_{\pi}$, 
the conditions (\ref{eq:7})-(\ref{eq:8}) are not fulfilled in $\Xi_{3/2}$ decays
and  the pictures
of the $m$- and $E$-bands shown in Figs 3c, and 3d
are very similar to those in $\Theta^+$ decays. At low momenta  
 the efficiency of the  $E$-selector is 100$\%$.
The $E$-bands start overlapping at $P_{\Xi}>$1.96 GeV.
Thus, the efficiency of the $E$-selector is
no worse than  84$\%$ (Fig.4).

\subsection{$\Xi_{3/2}(2070)\rightarrow \Sigma (1385) + K(498)$}
The $\Sigma (1385)$ decays strongly and exclusively at the primary
event vertex. The decay modes $\Sigma^{\pm\,0}\rightarrow \Lambda\pi^{\pm\,0}$
are dominant. Therefore, after the reconstruction of $\Lambda$,  
the use of the $m$- or $E$-selector  allows 
reconstruction of the  $\Sigma (1385)$ with 100$\%$ efficiency (see Fig 1b, Fig. 2b).

 Figs 3e and 3f  shows the $m$- and $E$-bands in $\Xi_{3/2}(2070)$ decays.
The $m$-bands  slightly overlap at low $P_{\Xi}$ and diverge at 
$P_{\Xi}>$ 7 GeV. The $E$-bands are totally separated. Thus, in that decay mode,
the $m$- and $E$-selectors  work with 100$\%$ efficiency (see Fig. 4).

\subsection{$\Xi_{3/2}(2070)$ or $\Xi_{3/2}(1860)$ ?}
The NA49 collaboration is provided  \cite{NA49} the  evidence for the existence of a narrow
$\Xi^-\pi^-$ baryon resonance with mass of 1.862$\pm$0.002 GeV. This state is considered
as a candidate for the exotic pentaquark state $\Xi_{3/2}$. The reported mass value  
 is much lower as predicted in \cite{DPP}. The last developments in the theory of pentaquark states did not  exclude the lower mass for the $\Xi_{3/2}$ \cite{polyakov}.
There are also arguments \cite{fischer} that the result of the NA49 collaboration perhaps is inconsistent with data collected over the past decades.

In Fig. 5 shown the $m$-bands and the efficiency. They are much the similar to those
in Fig. 3e and Fig. 4d. The picture of the $E$-bands is also similar to Fig. 3f.
Thus, the $E$-selector will not suppress the signal but suppress background at higher
masses.

\section{Conclusions}
Kinematics of the two-body decay, $R\rightarrow {\bf a}+{\bf b}$, has been analyzed in terms 
of the phase space $m$- and $E$-bands. On the basis of many examples,
in particular, the exotic anti-decuplet baryons (pentaquark states),
it has been demonstrated that for $m_{\bf a}>m_{\bf b}$
the selection rules $p_{\bf a}>p_{\bf b}$ and $E_{\bf a}>E_{\bf b}$ can be with a high
efficiency applied to reconstruct many resonances and to suppress backgrounds.

\vspace*{6mm}
\noindent
{\large \bf Acknowledgments}

\vspace*{3mm}
\noindent
The author is indebted to P. Bussey and A. Geise for a reading of the manuscript and comments,  and grateful to other colleagues from the ZEUS collaboration   
 for useful discussions. Special thanks are due to P. Ermolov  and A. Kubarovsky for discussions of 
preliminary results of the SVD collaboration.
This study is partially supported  by the RFBR under Grant no. 02-02-81023.

%\vspace*{10mm}

{}

\begin{thebibliography}{99}

\bibitem{ksks}
ZEUS Collaboration; S. Chekanov et al.,
Phys. Lett. {\bf B 578}, 33 (2004);\\ 
Preprint DESY-03-098 

\bibitem{kopbk} G.I Kopilov, {\it Basics of resonance kinematics} (in
Russian), Nauka, 1970;\\
E.Byckling and K.Kajantie, {\it Particle kinematics}, John Wiley$\&$Sons,
1972
\bibitem{pdg}
Particle Data Groupe, K. Hagiwara et al., {\it Review of particle physics}. Phys.
Rev. {\bf D 66}, 010001 (2002)
\bibitem{DPP}
D.Diakonov, V.Petrov and M.V.~Polyakov, Z.Phys. {\bf A\,359}, 305\,(1997).
\bibitem{5qfte1}
LEPS Collaboration, T. Nakano et al., Phys. Rev. Lett. {\bf 91}, 012002\,(2003);\\
CLAS Collaboration, S.\,Stepanyan et al., arXiv:hep-ex/0307018;\\
CLAS Collaboration,  V.~Kubarovsky {\it et al.}, arXiv:hep-ex/0311046;\\
SAPHIR Collaboration, J.\,Barth et all., Phys. Lett. {\bf B 572}, 127\,(2003);\\
 A.~E.~Asratayn, A.~G.~Dolgolenko and M.~A.~Kubantsev, arXiv:hep-ex/0309042;\\
DIANA Collaboration, V.V.\,Barmin et al., Phys.\,Atom.\,Nucl. {\bf 66}, 1715\,(2003);\\
HERMES Collaboration, A. Airapetian et al., arXiv:hep-ex/0312044 and DESY-03-213 
\bibitem{5qpte1}
ZEUS Collaboration., S. Chekanov, a talk at "DESY Forum: Pentaquarks at HERA",  DESY,
Hamburg, November, 25, 2003. http://webcast.desy.de

\bibitem{5qfte2}
SVD Collaboration., P.Ermolov, private communication.

\bibitem{NA49}
NA49 Collaboration, C. Alt et al., arXiv:hep-ex/0310014
\bibitem{polyakov}
M. Polyakov, a talk at "DESY Forum: Pentaquarks at HERA",  DESY,
Hamburg, November, 25, 2003. http://webcast.desy.de

\bibitem{fischer}
H.G. Fischer and S. Wenig, arXiv:hep-ex/0401014

\end{thebibliography}
\end{document}